\documentclass[useAMS,usenatbib]{mn2e}
\usepackage{psfig}
\usepackage{epsfig}
\usepackage{times}

\def\mjup{\mbox{$M_\textrm{\tiny Jup}$}} 
 
\def\msun{\mbox{$M_\odot$}}
\def\Msun{\mbox{$M_\odot$}}

\def\ga{\mathrel{\hbox{\rlap{\hbox{\lower4pt\hbox{$\sim$}}}{\raise2pt\hbox{$>$}}}}}
\def\la{\mathrel{\hbox{\rlap{\hbox{\lower4pt\hbox{$\sim$}}}{\raise2pt\hbox{$<$}}}}}

\begin{document}

\title[Limits on planetary-mass companions to vMa2]{The ``DODO''
  survey I: limits on ultra-cool substellar and planetary-mass    
companions to van Maanen's star (vMa\,2)}

\author[M.\,R. Burleigh et~al. ]
{M.\,R. Burleigh$^1$, F.\,J. Clarke$^2$, E. Hogan$^1$, C.\,S. Brinkworth$^3$, 
P. Bergeron$^4$, P. Dufour$^5$,  
\newauthor P.\,D. Dobbie$^6$, 
A.\,J. Levan$^7$, S.\,T. Hodgkin$^8$, D.\,W. Hoard$^3$ \&
S. Wachter$^3$ \\
$^1$ Department of Physics and Astronomy, University of Leicester,
University Rd., Leicester LE1 7RH, UK\\
$^2$ Department of Astrophysics, University of Oxford, Oxford OX1 3RH, UK\\
$^3$ Spitzer Science Center, California Institute of Technology, Pasadena, 
CA 91125, USA\\
$^4$ D\'epartment de Physique, Universit\'e de Montr\'eal, C.P.~6128, Succ.~Centre-Ville, Montr\'eal, Qu\'ebec, Canada, H3C 3J7\\
$^5$ Department of Astronomy and Steward Observatory, University of
Arizona, 933 North Cherry Avenue, Tucson, AZ\,85721\\
$^6$ Anglo-Australian Observatory, PO Box 296, Epping, NSW 1710, Australia\\
$^7$ Department of Physics, University of Warwick, Coventry CV4 7AL, UK\\
$^8$ Institute of Astronomy, University of Cambridge, Cambridge\\
}

\maketitle

\begin{abstract}

We report limits in the planetary-mass regime for   
companions around the nearest single white
dwarf to the Sun, van Maanen's star (vMa\,2), from deep $J$-band
imaging with Gemini North and {\it Spitzer} IRAC mid-IR photometry.  
%as the first results from the ``{\it
%  DODO}'' survey for planets in wide orbits around nearby white
%dwarfs.     
We find no resolved common proper motion companions to vMa\,2 at
separations from $3 - 45\arcsec$, at a limiting magnitude of 
$J\approx23$. Assuming a total age for the system of $4.1\pm1$\,Gyr, and
utilising the latest evolutionary models for substellar objects, this
limit is equivalent to companion masses $>7\pm1\,\mjup$
($T_{\rm eff}\approx300$\,K). Taking into account the likely 
orbital evolution of very low mass companions in the
post-main sequence phase, these $J$-band observations effectively 
survey orbits around the white dwarf {\it progenitor} from $3 -
50$\,AU. There is no flux excess detected in any of the 
complimentary {\it Spitzer} IRAC mid-IR filters
%photometric fluxes are 
%consistent with those predicted by a new 
We fit a 
DZ white dwarf model atmosphere to the optical $BVRI$, 2MASS $JHK$ and
IRAC photometry. The best solution gives $T_{\rm eff}=6030\pm240$K,
log~g$=8.10\pm0.04$ and, hence, $M= 0.633\pm0.022\Msun$. We then place a  
$3\,\sigma$ upper limit of $10\pm2\,\mjup$ 
on the mass of any unresolved companion in the $4.5\mu$m band.  

\end{abstract}


\section{Introduction}

Direct imaging of  
extra-solar planetary-mass companions to solar-type stars 
is complicated by problems of contrast and resolution. At
the time of writing, no planet has been directly imaged around a
solar-type star.
An alternative solution to these difficulties 
is to target intrinsically faint stars instead, such as white dwarfs. 
%For example, several groups are 
%searching for planetary mass companions to very young low mass stars 
%and brown dwarfs in nearby clusters and associations. 
%\citet{chauvina,chauvinb} have imaged a $\approx5\,\mjup$ 
%companion to a young brown dwarf in the 
%TW~Hya association (2MASSW\,J1207334$-$393254), although consideration
%of this object's likely formation process suggests 
%the pair should be better regarded as a brown dwarf binary. The masses
%of two other claimed directly imaged planets are currently too
%uncertain to confidentally classify them as such 
%(GQ~Lup\,B, \citealt{neuhauser}, \citealt{jansen}, \citealt{mcelwain}, 
%\citealt*{marois}; SCR\,1845$-$6357\,B 
%\citealt{SCR1845}), while the estimated mass of AB~Pic\,B
%($13.5 \pm 0.5 \mjup$ \citealt{abpic}) places it on the boundary
%with deuterium burning brown dwarfs.
Stellar evolution lends two huge 
advantages when searching for very faint companions: 
white dwarfs are up to $\sim10^4$ times fainter than their 
main sequence progenitors, 
and the orbits of any 
planetary-mass companions that 
lie outside the stellar envelopes during the giant phases 
will expand outwards as mass is lost 
from the central star, increasing the projected 
separation by a maximum factor $M_{\rm MS} / M_{\rm WD}$ \citep{jeans}. 
Thus, the problems of contrast and resolution are 
greatly reduced. The possible evolution of planetary systems 
in the post-main 
sequence phase is discussed in more detail by \citet{duncan}, 
\citet*{burleigh02}, \citet{Debes02} and \citet{villaver}. 

The direct detection of such low mass companions to white dwarfs
opens up the possibility for spectroscopic investigation of a
previously unobserved class 
of object: evolved low-mass 
brown dwarfs and planetary-mass gas giants as low in 
temperature as $T_{\rm eff}\approx300$\,K. 
In contrast, the directly imaged planetary-mass  
($\approx5\mjup$) companion to the brown dwarf 2MASSW\,J1207 has the
spectrum of a mid-L~dwarf \citep{chauvina,chauvinb}, because it is
still young ($\sim10^7$~years old). The coolest known brown dwarf, 
ULAS\,J$003402.77-005206.7$, has a temperature 
$600$K$< T_{\rm eff} < 700$K \citep{warren07} and a spectral type
T\,8.5. The letter ``Y'' has been
suggested for the next, cooler, spectral type \citep{Kirkpatrick},
which might include evolved planetary-mass companions to white
dwarfs. Alternatively, if no obvious spectral change triggers the use
of a new letter, then the T classification will need to be extended
beyond T\,8.5. 
%(see \citealt{Leggett07} for a discussion). 

%Planetary systems around white dwarfs are
%of interest in themselves as part of the broader topic of
%``comparative planetology'', i.e. the study of planetary systems in a
%variety of environments, and for providing information on the final
%evolutionary stages of solar systems. In addition, the white
%dwarfs most suitable for searches for planetary-mass companions are
%descended from early-type stars (spectral types mid- 
%and late-B, A and early-F) which are
%not usually the subject of radial velocity searches. By searching for
%planetary-mass objects around these white dwarfs, we provide information on the
%frequency and mass distribution 
%of planetary systems around their relatively massive
%progenitors. 

%new para on spectroscopic follow-up, comparative planetology and 
% usefulness for studying planetary companions to progenitor stars

%not new idea, older searches (summarize other searches later) & pulsations

The idea of using white dwarfs to find intrinsically faint, low mass 
companions is not new. \citet{probst83} and \citet{becklin88} used 
the low luminosity of white dwarfs to search for 
brown dwarf companions as near-infrared photometric excesses, and indeed  
the latter achieved success with GD\,165\,B (L\,4). 
More recently, \citet{fbz05} have conducted a 
comprehensive search for brown dwarf companions to 
several hundred white 
dwarfs but detected only one new pair (GD\,1400, L\,$6-7$, 
\citealt{GD1400}, \citealt{dobbie05}), while \citet{maxted06} and
\citet{wd0137b} have detected a close L\,8  
companion to the white dwarf WD\,0137$-$349. No other detached substellar
companions to white dwarfs are known.

The most intriguing circumstantial 
evidence for the existence of old planetary systems 
that have survived to the final stage of stellar evolution 
comes from the discovery of metal-rich 
circumstellar dust and gas disks around a growing number of 
white dwarfs \citep{zuckerman87,becklin05,gd362,gd56,sdss1228,vh07,
Jura07,WD1150}. 
%\citet{zuckerman87} 
%first identified the large infrared excess in 
%G\,29$-$38, which was subsequently shown to be due to circumstellar dust 
%rather than a substellar companion \citep*{tokunaga,chary}. 
{\it Spitzer} 
mid-infrared spectroscopy has now revealed that the dust   
disks are composed largely 
of silicates \citep{Reach,JuraGD362}. 
The favoured explanation for the origin of this 
material is the tidal disruption of an asteroid that has wandered into the 
Roche radius of the white dwarf \citep{Jura1},  
%Since the red giant progenitor 
%of the white dwarf would have had a radius $\sim1000$R$_\odot$, the 
%asteroid must have been forced inward subsequently, 
perhaps through interaction with 
planets in a solar system that has become destabilized in the wake of 
the planetary nebula phase \citep{Debes02}. 
%Although no planetary companions 
%to G\,29$-$38 have been found yet \citep*{Debes1}, the number of white dwarfs 
%known to possess circumstellar debris disks is growing fast 
%\citep[e.g.~][]{becklin05,gd362,gd56,sdss1228,vh07}. Indeed, 
\citet{Jura2} 
suggests that at least 7\% of white dwarfs possess asteroid belts. If that 
is indeed the case, it is likely that at least this number of white dwarfs 
also possess planetary systems. 

The recent discovery of a $M {\rm sin} i = 3.2\mjup$ planet in a
$1.7$\,AU orbit around the
extreme horizontal branch star V391\,Pegasi by \citet{V391Peg} proves that such
objects can survive red giant branch evolution, strongly
suggesting that it is simply a matter of time before a planet is discovered
around a white dwarf.

%The opening of access to 8m class telescopes, together with new predictions 
%for the evolution and expected luminosities of mature massive Jovian planets 
%(e.g.~\citealt{Burrows97}) led 
\citet{burleigh02}  made predictions concerning the 
likely near-infrared brightness of putative resolved, 
giant planetary companions to 
nearby white dwarfs, based on their likely total ages 
%(white dwarf cooling age plus 
%main sequence lifetime) 
and distances. In 2002 
we initiated a programme to search for wide, spatially resolved very low mass 
common proper motion companions to white dwarfs via direct imaging. 
In particular, we aim to find substellar companions with masses 
greater than a few~$\,\mjup$ around white dwarfs within $\approx20$~pc
of the Sun. 
Such companions are expected to have near-IR 
magnitudes brighter than $J\sim23.5$, 
commensurate with the expected sensitivity of an 
8m telescope in a one hour exposure. 

\begin{sloppypar}  
We christened our project {\it ``DODO''} 
-- {\it D}egenerate {\it O}bjects around 
{\it D}egenerate {\it O}bjects. Preliminary results and  
progress reports have been published elsewhere \citep*{dodoa,dodob,dodoc}. 
Here, we report our results for the nearest single white dwarf to the Sun 
in our sample, van Maanen's star (vMa\,2, WD\,0046$+$051, d$\,= 4.41$\,pc,
\citealt{Hipparcos}). 
We combine two epochs of 
deep, ground-based 
$J$-band imaging from the 8m Gemini North telescope 
with {\it Spitzer} mid-infrared photometry to place limits on 
the masses and temperatures of any common proper motion
and unresolved ultra-cool substellar
and planetary-mass companions. 
%Our target selection criteria gave us a sample of 40 suitable white dwarfs, 
%which we have been observing using Gemini North $+$ {\it NIRI}, 
%Gemini South $+$ {\it Flamingos} (2002 only), 
%and the VLT $+$ {\it ISAAC} (2003 onwards). Note that we 
%do not employ adaptive optics, as we are primarily concerned with searching 
%for companions in the wide orbits predicted by stellar evolution models, 
%outside the white dwarf's point spread function. In this paper we report 
%on the results of our survey for common proper motion giant planetary 
%companions to a subset of 25 northern and equatorial white dwarfs. 
\end{sloppypar}

\subsection{vMa\,2 and previous infrared observations}

vMa\,2 was discovered serendipitously by \citet{vMa2} in a survey for
common proper motion companions to an unrelated star,
HD\,4628. It has a cool, helium-rich atmosphere and strong resonance
Ca\,II H \& K lines in its optical spectrum, and is classified as a DZ
white dwarf. The heavy elements are expected to sink below the
photosphere on a timescale much shorter than the white dwarf cooling
time. Their presence has been explained previously 
in terms of episodic accretion
from the interstellar medium \citep{Dupuis92}, but this
scenario requires the accretion rate of hydrogen to be at least two
orders of magnitude lower than the metals (\citealt{Wolff02,
  Dupuis93, Dufour}). Alternatively, the presence of metals in DZs might be
explained from accretion of cometary material or tidally disrupted
asteroids and planets. However, no DZ has been identified with an
infrared excess due to dust emission, possibly because the diffusion timescale
for heavy elements in cool white dwarf helium-rich atmospheres is
long and the material may have been accreted as much as
$\sim10^6$~years ago. Nonetheless, DZ white dwarfs are a speculative 
candidate for hosting  old planetary systems and vMa\,2, as the
nearest single white dwarf to the Sun, is an ideal target for such a
search. 

%Later, \citet{vMa2b} noted it to be ``by far
%the faintest F-type star known at the present time''. Of course, the
%true nature of white dwarfs was not known to van Maanen in 1919. 
Through analysis of {\it Hipparcos} data, \citet{Makarov} claimed to
have detected astrometrically a $0.06 \pm 0.02\msun$ 
substellar companion to vMa\,2, with an orbital period of 1.57~years
and a maximum separation on the sky of $0.3\arcsec$. \citet*{fbm}
carried out a search for this companion by direct imaging 
with adaptive optics in the mid-infrared $L'$ band. They also looked
for an unresolved companion as a near- and mid-infrared photometric
excess in ground-based and {\it ISO} photometry. 
They refuted the existence of Makarov's companion, and of any excess
emission due to dust, and placed
a limit of $T_{\rm eff} \la 500$\,K  
on the temperature of any putative substellar companion.

\begin{table*}
\caption{Adopted parameters for van Maanen's star}
\begin{center}
\begin{tabular}{cccccccccc}
\hline
$\mu\,^a$ & $\theta\,^a$ & d$\,^a$ & $T_{\rm eff}\,^b$ & log~$g\,^b$ & $M_{\rm WD}\,^b$ & 
$t_{\rm WD}\,^c$ 
& $M_{\rm MS}\,^d$ & $t_{\rm MS}\,^e$ & $t_{\rm total}$ \\
(mas/yr) & (mas/yr) & (pc) & (K) & & ($\msun$) & (Gyr) & ($\msun$) & (Gyr) 
& (Gyr) \\
\hline
\hline
1231.72 & $-2707.67$ & 4.41 & 6030 (240) & 8.10 (0.04) & 0.633 (0.022) &
3.17 (0.29) & 2.6 & 0.9 & 4.1 \\
\hline
\end{tabular}
\end{center}
$^a$~Hipparcos measurements, $^b$~from our new fit to the optical,
near-IR and IRAC mid-IR photometry (see section~4.2),  
$^c$~estimated using evolutionary models appropriate for
He-atmosphere white dwarfs with C/O core compositions, see
http://www.astro.umontreal.ca/~bergeron/CollingModels/, 
$^d$~derived from the 
initial-final mass relation of \citet{Dobbie06}, $^e$~\citet{Wood}
\end{table*}

\section{Observations \& Data Reduction}

\subsection{Gemini NIRI {\bf $J$}-band imaging}

Two deep $J$~band images centred on vMa\,2 
were acquired using Gemini North and the Near-Infrared Imager 
(NIRI, \citealt{NIRI}) with the f/6 camera 
($0.117\arcsec$ pixel$^{-1}$, field of view $120\arcsec \times 120\arcsec$)
on 2004 December 25th and 2005 August 25th, in queue scheduled mode
under programmes GN-2004B-Q-23 and GN-2005B-Q-19. 
The images were acquired with 60~s integrations at each position in a  
54 point dither pattern,  
for total exposure times of 54 and 75~minutes respectively. The 
measured FWHM of stars in the images 
was $0.39\arcsec$ and $0.69\arcsec$ in the two epochs
respectively.   
The data were reduced in the standard manner using the Gemini NIRI package in 
IRAF. A more detailed description of our 
data reduction method will be described in a later publication \citep{Hogan}. 
We note here that both these images were degraded by intermittent 
60\,Hz electronic 
pick-up noise, which manifests itself as a diagonal herringbone
pattern \citep{NIRI}. 
%This pattern is not present in the lab and has 
%been eliminated at 
%times on the telescope, indicating that it arises from the telescope 
%environment and not from the NIRI electronics. This noise has not been 
%removed from these data.

%pattern as a 
%variety of diagonal, vertical and horizontal  
%patterns, which are repeated in each quadrant of the detector.  
%These patterns are not present in the lab and have been eliminated at 
%times on the telescope, indicating that they arise from the telescope 
%environment and not from the NIRI electronics. We developed a task to aid 
%removal of these noise sources by collapsing each quadrant along lines and 
%columns, and subtracting the median value from each line and column.

\subsection{{\it Spitzer} IRAC mid-infrared observations}

Mid-infrared observations of vMa\,2 were obtained with the Infrared Array 
Camera (IRAC, \citealt{IRAC}) on the {\it Spitzer} Space Telscope 
\citep{Spitzer} on 2004 July 4th as part of Guaranteed Time programme 
PID00033 (PI: G. Fazio). IRAC obtains images 
in four channels with central wavelengths of $3.6\mu$m, 
$4.5\mu$m, $5.8\mu$m and $8.0\mu$m. 
We downloaded the {\it Spitzer} IRAC 
data from the public archive as Basic Calibrated
Data (BCD), which have been reduced and flux calibrated with the S14
version of the IRAC pipeline. The BCD images were corrected in two different
ways as a check on our reduction method. Firstly, we corrected the
individual images for array-location dependence, using the correction images
provided by the Spitzer Science Center (SSC), before carrying out the
photometry on the individual corrected BCDs. Secondly, we
corrected the BCDs for array-location-dependence using the correction images
that had been divided by the pixel-distortion images (also provided by the
SSC). We combined these corrected BCDs with dual-outlier rejection to
produce a mosaicked image for each channel using the SSC mosaicking
software, MOPEX \citep{MOPEX}. 
By dividing the array-location-dependence images by the
pixel distortion images, we avoid doubly correcting for pixel distortion
when combining the BCDs with MOPEX. Both of the reduction methods gave the
same results to well within uncertainties.
%We have downloaded these data from the public archive as Basic Calibrated 
%Data (BCD), which have been 
%reduced and flux calibrated with the S14 version of the IRAC
%pipeline. These BCD images 
%were then corrected for array-location-dependency using 
%the correction frames provided by the {\it Spitzer} Science Center, 
%and combined 
%with dual-outlier rejection to produce a mosaicked image for each channel 
%using the {\it Spitzer} Science Center mosaicking software, MOPEX \citep{MOPEX}.

\section{Data Analysis}

\subsection{{\bf $J$}-band images}

%The positions of all objects in the final stacked image must be accurately 
%determined before precise proper motions can be measured.
Each Gemini NIRI $J$-band image was set 
%roughly (within $\approx 1 -- 2\arcsec$) 
to the World 
Co-ordinate System using stars in the field with good 2MASS $J$-band 
detections \citep{2MASS}.  
We used the SExtractor task to detect all point sources in the 
final stacked image with a flux $\ge3\,\sigma$ above deviations in 
the local background, 
and computed the motions of these objects in RA and Dec between the two 
epochs (Figure~1). We also show in Figure~1 the $1\,\sigma$ and $3\,\sigma$
scatter of the distribution of all the measurements (excluding the
white dwarf). 
%We do not attempt to measure the proper motions of objects 
%indicated by SExtractor to have an ellipticity $e > 0.5$, which we
%assume are extended galaxies.  

The instrumental magnitude of each detected source was corrected to 
apparent magnitude using stars in the field with 2MASS $J$-band 
fluxes \citep{2MASS}. 
To calculate the $3\,\sigma$ completeness limit of these data, we inserted  
a total of 10,000 fake stars into each image at a variety of 
magnitudes from $J = 19 - 24$, and attempted to recover them using 
SExtractor. Figure~2 shows the fraction of implanted 
stars recovered as a function 
of their brightness at each epoch. The number of implanted 
stars recovered is always less than 100\% of those injected  
as some stars are lost behind or 
within the wings of other real or implanted stars. 
The fake stars were inserted at the same positions in {\it both} images. 
However, each implanted star may not necessarily 
be recovered in both images, especially towards the completeness 
limit, in which case it would not be available for a proper motion 
measurement. The curve with the star markers takes this into account. 

\begin{figure}
\centering
\caption{Proper motions of all point sources detected 
in both the NIRI $J$-band images,
between 2004 December 25th and 2005 August 25th. The white dwarf's
motion is clearly discriminated from all other objects in the
field. The dashed circles represent the $1\,\sigma$ and $3\,\sigma$ 
scatter of the distribution of the proper motions, centred on the 
white dwarf for clarity.}
\includegraphics[bb=42 110 576 655,angle=270,scale=0.45]{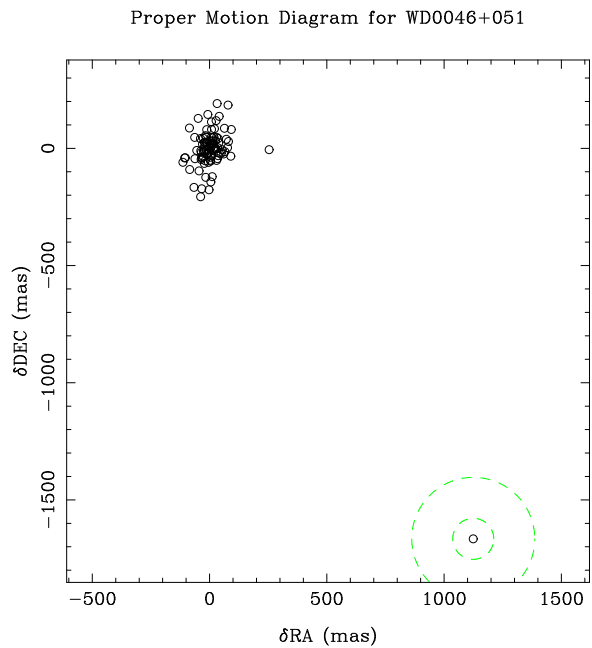}
\end{figure}

\begin{figure}
\centering
\caption{The completeness limit 
of each NIRI observation is determined by injecting
  fake stars into each image and recording the fraction recovered by
  our source detection method against magnitude, as described in the
  text. We also show the fraction recovered in {\it both} images.}
\includegraphics[bb=42 110 576 655,angle=270,scale=0.45]{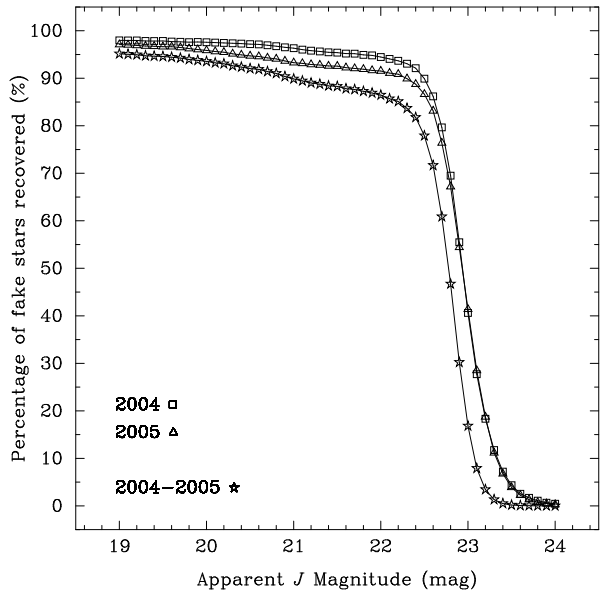}
\end{figure}

%\mbox{\includegraphics[angle=270,scale=0.5]{WD0046+051_sensitivity.ps}}
%\caption{}
%\end{figure*}
%\end{center}

%\begin{center}
%\begin{figure}
%\mbox{\includegraphics[angle=270,scale=0.5]{WD0046+051_pm.ps}}
%\caption{}
%\end{figure}
%\end{center}

\subsection{{\it Spitzer} IRAC mid-infrared photometry}

We performed aperture photometry on the mosaicked IRAC images using 
standard IRAF tools. 
A full description is given in \citet{Carolyn}. Briefly, 
we adopted an aperture size of 2.5~pixels, and 
subsequently corrected to an aperture size of 10~pixels.  
using 
%corrections derived from sources in the field. 
%The values of the aperture 
corrections of 1.190, 1.199, 1.211 and 1.295 for channels $1-4$ respectively. 
%From experience we found that the point response function (PRF) for the IRAC 
%cameras is not well-modelled and we obtain more consistent results using 
%aperture photometry than via PRF fitting. Sky subtraction was tested in 
%three ways: (1) using a sky annulus around the target; (2) using a sky
%region 
%a short distance from the target; and (3) subtracting a median sky from the 
%mosaic before performing the photometry. The three methods were found to be 
%consistent with each other. 
The IRAC fluxes of vMa\,2 are given in Table~2. 
The errors consist of the quadratic addition of the 
standard deviation of the fluxes in individual frames plus the     
systematic error on the calibration of each filter as determined by 
\citet{iraccal}.
%We note that these fluxes are consistent with those from a second IRAC
%observation of vMa2 in 2006. 

\begin{table}
\caption{Predicted and observed {\it Spitzer} IRAC fluxes}
\begin{center}
\begin{tabular}{ccc}
\hline
 & Predicted & Observed \\
$\lambda_0$ & $F_\nu$ & $F_\nu$ \\
($\mu$m) & (mJy) & (mJy) \\
\hline
\hline
3.56 & 7.991 & $8.030\pm0.262$ \\
4.51 & 5.370 & $5.451\pm0.170$ \\
5.76 & 3.551 & $3.611\pm0.126$ \\
7.96 & 2.038 & $2.092\pm0.089$ \\
\hline
\end{tabular}
\end{center}
\end{table}

\section{Results}

\subsection{Limits on common proper motion companions}

Figure~1 shows the proper motions of all point sources detected in the NIRI 
images, between the two epochs. The white dwarf is clearly discriminated from 
everything else in the field, i.e.~we do not detect any common proper motion 
companions to vMa\,2. We recover 90\% of fake stars injected into both images 
at $J\approx20.9$~mags, 
and 50\% at $J\approx22.7$~mags (Figure~2). These sensitivity 
estimates can be translated into limits on the mass and temperature of any 
resolved companions, using appropriate evolutionary models for 
substellar objects. We adopt the COND models of 
\citet{Baraffe03}, which give the temperature and luminosity of substellar 
objects evolving in isolation, i.e.~we assume no insolation from the white 
dwarf, and neglect any possible heating during previous evolutionary phases.
The age of vMa\,2 can be calculated by combining the white dwarf cooling age 
($3.17\pm0.29$\,Gyr, see below) 
with an estimate of the main sequence lifetime. Using the 
initial-final mass relation of \citet{Dobbie06}, we estimate that the 
progenitor of vMa\,2 ($M_{\rm WD} = 0.633\pm0.022\,\msun$, see below) 
had a mass 
$= 2.6\msun$. We estimate the lifetime of such a main sequence star as 
$\approx0.9$\,Gyr, 
from the relationship of \citet{Wood}. Thus, we adopt a  
total age for vMa\,2 of 4.1\,Gyr. At this age, and  
taking the {\it Hipparcos}-derived distance to vMa\,2 of 4.41~pc, the 90\% 
completeness limit ($J\approx20.9$~mags) translates to a 
limit on the mass of resolved companions of 
$\approx0.009\msun$ ($\approx9\mjup$, $T_{\rm eff}\approx320$\,K), and the 
50\% limit ($J\approx22.7$~mags) translates to a limiting 
mass of $\approx0.007\msun$
($\approx7\mjup$, $T_{\rm eff}\approx280$\,K). 

These mass and temperature estimates are 
of course model-dependent. For example, 
the white dwarf initial-final mass relation
is empirical and the subject of several current studies
(e.g.~\citealt{Dobbie06}; \citealt{Kalirai05}; \citealt{Ferrario05}),    
white dwarf evolutionary models are constantly being revised and
improved, and  
the lifetimes of field main sequence stars 
are notoriously difficult to estimate. 
%Most importantly, we assume that the 
%evolutionary models for mature gas giants are correct, although they
%are untested by observation. 
At ages $>1$\,Gyr the evolutionary models for substellar objects predict
that they cool very slowly, and are relatively
insensitive to age. Therefore, even if we adopt a very conservative 
error of $\pm25\%$ ($\pm1$\,Gyr) on the total 
age of vMa\,2 then  
the errors on the masses and temperatures of putative companions are
small:  
$9 \pm 1\mjup$ ($300 \pm 20$\,K) at the 90\% completeness limit 
and $7 \pm 1 \mjup$ (also, $300 \pm 20$\,K) at the 50\% completeness limit.  

The minimum projected physical separation 
surveyed is given by the distance from the 
centre of its point spread function where the white dwarf 
is indistinguishable from the background. 
We estimate this as $\approx3\arcsec$, 
equivalent to $\approx13$\,AU at the distance of vMa\,2. The maximum 
projected physical separation 
surveyed is given by the field of view covered by both NIRI
images. Although the field of view of the NIRI f/6 camera is nominally 
$120\arcsec \times 120 \arcsec$, the dither pattern used during the
observations restricts the useable field of view to   
a maximum radius of $45\arcsec$, equivalent to $\approx200$\,AU. 
As mass is lost from the central star during the post main sequence phase, 
the projected star--planet separation for planets outside the red
giant envelope will increase 
by a maximum factor $M_{\rm MS} / M_{\rm WD} = 2.6\,\msun / 0.63\,\msun
\approx 4$ \citep{jeans}. From this 
we estimate the range of orbital radii surveyed around the white dwarf 
{\it progenitor} as $3 - 50$\,AU. 

% First deal with J-band images: limits on Jovian masses and temperatures 
%of putative companions andlikely errors, area surveyed

\subsection{Limits on unresolved companions through {\it Spitzer} 
mid-infrared photometric excesses}

%\begin{figure}
%\centering
%\caption{Optical and infra-red spectral energy distribution of vMa\,2
%  from $0.3\mu$m to $9.0\mu$m. The observed 2MASS $J\,H\,K_s$ and {\it
%  Spitzer} IRAC fluxes are shown by error bars. The DZ white dwarf 
%model monochromatic flux is shown as a solid line, and the fluxes
%  averaged over the filter passbands are indicated by filled
%  circles. The cross represents the expected flux of the white dwarf
%  plus an unresolved, $12\mjup$ companion.}
%\psfig{file=all.ps,angle=270,width=8.cm}
%\end{figure}

\begin{figure}
\centering
\caption{
Mid infra-red spectral energy distribution of vMa\,2
  from $3\mu$m to $10\mu$m. The observed {\it
  Spitzer} IRAC fluxes are shown by error bars. The DZ white dwarf 
model monochromatic flux is shown as a solid line, and the model fluxes
  averaged over the filter passbands are indicated by filled
  circles. Also shown for comparison are the $3.76\mu$m $L'$ band flux from
  \citet{fbm} and the $ISO$ $6.75\mu$m LW2 flux from \citet{chary}.}
\psfig{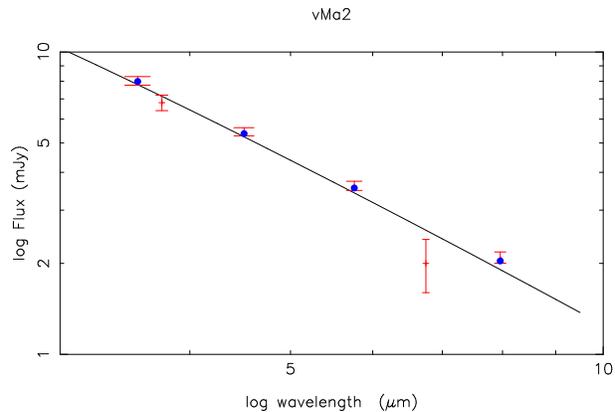}
\end{figure}

In order to place limits on the masses and temperatures of unresolved 
companions to vMa\,2, we 
compare the {\it Spitzer} IRAC photometry to an appropriate 
white dwarf model atmosphere plus COND models calculated for the IRAC 
bandpasses (Figure~3, Table~2). 
%We found that the model determined by
%\citet{Dufour} underpredicted the observed IRAC fluxes in all
%filters by a similar factor. 
%The difference could not be accounted for by an unresolved
%substellar companion (the COND models 
%predict that mature, substellar objects 
%will give a far greater flux excess 
%in the $4.5\mu$m band than at $5.8\mu$m or $8.0\mu$m)  
% or by the addition of a cool
%blackbody, such as might be expected if a debris disk was
%present. Instead, 
Firstly, 
we fitted a grid of DZ white dwarf model atmospheres to the optical (Bessel)
$BVRI$, near-IR 2MASS $JHK$ and the IRAC fluxes. The best fit solution
gives $T_{\rm eff} = 6030\pm240$~K and, when combined with the
parallax, log~$g = 8.10\pm0.04$. Within the formal uncertainties, 
these values are consistent with those determined by 
\citet{Dufour}, which were based on fitting the same $BVRI$ data but a
different set of CIT $JHK$ photometry, and did not include the IRAC
photometry. The IRAC fluxes now give us a much better leverage on the
atmospheric parameters. We then folded this   
model through the IRAC filter passbands to predict the IRAC fluxes (Table~2).
Our new predicted model is consistent with
the observed fluxes in {\it all} optical and infrared bandpasses,
within uncertainties. From evolutionary models, the best-fit
parameters lead to estimates for the
white dwarf mass of $0.633\pm0.022\Msun$ and cooling age
$3.17\pm0.29$~Gyr. 
%The total age of the system ($4.1$~Gyr)
%is the same (since the main sequence progenitor's mass is lower,
%hence its lifetime is longer, see Table~1). 
%We then folded this  
%model through the IRAC filter passbands to predict the IRAC fluxes (Table~2). 
%To these 
%predicted fluxes we have added COND models, again assuming an age for 
%vMa\,2 of $4.0\pm1.0$\,Gyr 
%and adopting a distance of $4.41$\,pc.
%until the predicted 
%IRAC fluxes are $3\,\sigma$ in excess of the white dwarf model. 

By adding COND models to the predicted IRAC $4.5\mu$m flux,
again assuming an age for vMa\,2 of $4.1\pm1.0$\,Gyr 
and adopting a distance of $4.41$\,pc, 
we find that in this band {\it Spitzer} 
could have detected a $10\pm2\mjup$, $T_{\rm eff} \approx 280 -
360$\,K planet at the $3\sigma$ level. 

% Then Spitzer: the WD model, predicted mags, agreement with observation, 
% and addition of planet models (3 sigma limits)
% Finally, given size of AGB star, will have efectively surveyed whole 
% are where planets expected, to lower T than Farihi 

\section{Summary}

We have placed limits on ultra-cool 
substellar and planetary-mass
objects around the nearest 
single white dwarf, vMa\,2, through a search for common proper motion 
companions and mid-infrared photometric excesses. 
The red giant progenitor to 
vMa\,2 had a maximum radius of $\sim1000R_\odot$ ($4.6$\,AU, 
\citealt*{Hurley}). Therefore, taking into account 
the post main sequence evolution of the orbits of any companions, 
our $J$-band images have covered all orbits of  
substellar objects that originally lay beyond the maximum extent of 
the red giant envelope, up to 50\,AU distance.  
The complimentary {\it 
Spitzer} IRAC photometry places slightly higher limits on the masses 
and temperatures of any spatially unresolved 
substellar companion whose orbit fortuitously 
lies along the line of sight to vMa\,2, or that currently has 
an orbit within $\approx13$\,AU of the white dwarf.  
%Putative very low mass
%brown dwarf or giant planet companions in such a close orbit would
%presumably have had to survive
These limits are significantly lower than those reported by previous
studies, e.g. \citet{fbm} and \citet{fbz05}.

A variety of searches for planetary companions to white dwarfs are currently 
underway (e.g.~\citealt{dodob}; \citealt{Debes1}; 
%\citealt{MullallyA}; 
\citealt{MullallyB}).  
Further, extensive results from the {\it DODO} survey are 
in preparation \citep{Hogan}. 

\section{Acknowledgments}

MB acknowledges receipt of a STFC Advanced Fellowship. EH
acknowledges the support of a STFC Postgraduate Studentship.  
This work is supported in part by the NSERC Canada. PB is a Cottrell
Scholar of Research Corporation. Based
in part on observations obtained at the Gemini Observatory. 
%which is
%operated by the Association of Universities for Research in Astronomy,
%Inc., under a cooperative agreement with the NSF on behalf of the
%Gemini partnership: the National Science Foundation (United States),
%the Particle Physics and Astronomy Research Council (United Kingdom),
%the National Research Council (Canada), CONICYT (Chile), the
%Australian Research Council (Australia), CNPq (Brazil) and CONICET
%(Argentina). Also based in part on observations made with the Spitzer Space
%Telescope, which is operated by the Jet Propulsion Laboratory,
%California Institute of Technology under a contract with NASA.  This
%publication makes use of data products from the Two Micron All
%Sky Survey, which is a joint project of the University of
%Massachusetts and the Infrared Processing and Analysis
%Center/California Institute of Technology, funded by the National
%Aeronautics and Space Administration and the National Science
%Foundation. 
We thank Isabelle Baraffe and Jay Farihi for their
enlightening thoughts and comments. 

\bibliographystyle{mn}
\bibliography{wdplanets,browndwarfs,mbu}

\end{document}